\documentclass[aps,amsmath,amssymb,pre,twocolumn,showpacs,superscriptaddress]{revtex4-2}
\usepackage{enumitem}
\usepackage{graphicx}
\usepackage{natbib}
\usepackage{epstopdf}
\usepackage{amssymb}
\usepackage{url}
\usepackage{amsmath,physics}
\usepackage{color}
\usepackage{comment}
\usepackage{soul}
\setstcolor{red}
\setulcolor{blue}
\setul{}{0.2ex}
\numberwithin{equation}{section}

\begin{document}
\date{\today}
\title{Periodic and non-periodic brainwaves emerging via random syncronization of closed loops of firing neurons.}
\author{P. Mazzetti}
\author{A. Carbone}
\affiliation{Politecnico di Torino, Italy}
\begin{abstract}
Periodic  and nonperiodic  components of electrophysiological signals are modelled in terms of syncronized sequences of  closed loops of firing neurons correlated in Markov chains. 
Single closed loops of firing neurons  reproduce  fundamental  and  harmonic components, appearing as lines in the  power spectra at frequencies ranging  about from $0.5 Hz$  to $ 100 Hz$.  
Further interesting features of the brainwave signals emerge by considering multiple syncronized sequences of closed loops. In particular, we show that the fluctuations of the number of syncronized loops leads to the onset of  broadband power spectral components.
 By effect of the fluctuations of the number of synchronized loops and the emergence of the related broadband component,  highly  distorted waveform and nonstationarity of the signal are observed,  consistently with  empirical  EEG and MEG signals. 
The analytical relationships of the periodic and aperiodic components are evaluated  by using  typical  firing neuron pulse amplitudes and durations.
\end{abstract}
\maketitle

\section{Introduction}
\label{Introduction}
Periodic components of brain signals and their frequency bands (\textit{delta} ($1-3 \mathrm{Hz}$), \textit{theta} ($4-8 \mathrm{Hz})$, \textit{alpha} ($9-12 \mathrm{Hz})$, \textit{beta} $(12-30 \mathrm{~Hz})$, \textit{gamma} $(>30 \mathrm{~Hz})$) are central to neuroscience basic research and clinical protocols \cite{daSilva2013EEG,Hirsch2010Atlas,Baillet2017Magneto}. Aperiodic components, initially disregarded in comparison to periodic ones as considered just background noise, represent a significant part of  signals. They manifest with power spectral densities varying approximately as $1/f^\beta$ and have been related to brain critical states \cite{deArcangelis2010Learning,Tagliazucchi2012Criticality}. Recent studies have suggested that simultaneous changes of aperiodic and periodic brainwave components can underpin changes in functional and behavioural features,  with the broadband components modulated by task performance and correlated with neuronal spiking activity. Synchronization between different neuronal groups  may also manifest within arrhythmic brain activity with no apparent periodicity  \cite{Buzsaki2004Neuronal, Canolty2006High,He2010The,Becker2018Alpha, Buzsaki2012Mechanisms,Wairagkar2021Dynamics,Brady2022Periodic,Cabral2022Metastable}. To keep pace with these findings, algorithms  are being developed  to the purpose of  breaking complex electrophysiological signals down and transferring scientific findings into clinical practices  \cite{Donoghue2020Parameterizing,Gerster2022Separating}, an  issue of  is increasingly relevant  to the development of brain–machine interfaces \cite{Zhang2020Electronic}. Despite remarkable advances in the interpreting and quantifying neurological signals, several problems mainly related to the dynamics of brain  at various scales still remain unsolved \cite{Goodfellow2022what}.
\par
In this work, a unified framework  to quantify  periodic and aperiodic power spectral  components of electro- and magneto-encephalograms  is developed based  on  a \textit{Markov chain} description. The power spectral density is estimated in terms of time-sequences according to a statistical approach originally pioneered in the communication and information theory context \cite{Huggins1957Signal,Barnard1964On,Mazzetti1978Spectral,Galko1981Spectral,Bilardi1983Spectral,Mazzetti2017Harmonic,Centers2021Power}.
Line spectral components are generated by \textit{Markov matrices} corresponding to closed loop sequences of firing neurons characterized by a set of heterogeneous states. 
The oscillatory frequencies, observed in the EEG and MEG  from 0.5 Hz  to 100 Hz, can be reproduced by closed loops involving a few hundred neurons down to few neurons with  firing intervals of the order of few milliseconds. 
The closed loop operates as an electric circuit where the neuron behaves as a rectifying diode, producing unidirectional currents, the synapses act as dissipative elements and the ionic currents  as current generators. The formation of closed loops  dissipate the excess energy accumulated in an active region of the brain by the local increase of circulating blood. 
A realistic description of the neurological signals require bunches of synchronized sequences of  closed loops of firing neurons spontaneously formed in regions of high  density, where neurons are connected to thousands other  neurons nearby, that, if in a critical state, may be simultaneously fired.  
\par
The  impulse generated by the  bunch of syncronized sequence is represented by a Gaussian time function. The width of the Gaussian is related to the duration of the impulse emitted by the bunch and, in turn, to the characteristic cut-off  filtering out the harmonic components of the power spectrum at high frequency. The  amplitude of the Gaussian depends on the number of syncronized loops $N_{i}$. By taking into account the fluctuations of $N_{i}$  in the Markov chain model, a mixed power spectrum is obtained where  lines and continuous spectral components co-exist.  It is shown that the broadband component causes  the line amplitude  to change, while their frequency keeps on unchanged being only dependent on the reciprocal duration of the loops,  and the distortion of the signal consistently with what is observed in the empirical EEG and MEG records. This distortion affects the lowest rather than the highest frequency components: thus \textit{delta waves} with  harmonics at frequencies lower than the  cut-off are more distorted than \textit{gamma waves}. In general the waves reported in electrophysiological  graphs are  at least distorted by second harmonic components, resulting in  asymmetric triangular form of the wave.
\par
Tha manuscript is organized as follows. 
In Sect.2  the general expression of  the  power spectral density of a sequence of Markov correlated events is recalled. The conditions required for the onset  of open or closed   loops in the framework of the Markov chain description are also provided. Then, the approach is extended to an arbitrary number of synchronized closed loops, whose fluctuations causes the emergence of a broadband noise component. In Sect. 3,   the proposed analitycal framework is used  as descriptive background  to reproduce brainwaves signal features in relation to the mixed power spectrum. In Sect. 4, the proposed models and results are discussed, conclusions and suggestions for future work are also provided.
\section{Mathematical Framework}
In this section, the  mathematical background for the calculation of  the power spectral density  of a sequence of events correlated in a Markov chain  and  conditions to yield  a closed loop are recalled.   The amplitude of the periodic and aperiodic components are  derived for a single loop  in subsection (A) and for an arbitrary number of synchronized closed loops   in subsection(B).
\par
Consider a neuron,  in a state labelled  $\alpha_1 $, making a firing to a neuron, in a state labelled  $\alpha_2$,  and so on in a sequence  $ \alpha_1, \alpha_2, ..., \alpha_N $  of $N$  states. Let $n_1, n_2, \ldots, n_N$ indicate the numbers of  neurons respectively  in the states $\alpha_1,\alpha_2\ldots  \alpha_N $.  The firings of  $n_1$ synchronized neurons result in the subsequent firing of $ n_2$ neurons, yielding a total of $n_1 \cdot n_2$ synchronized  neurons, and so on. Hence, the number of synchronized neurons may reach a value of the order of several thousands in a relatively short time.  For the sake of the example if each of the $ n_1, n_2,... $ produces a pair of simultaneously firing neurons  with firing characteristic time of $5 ms$, a total number $N_{i}$ of  about $ 10^7$ synchronized neurons would be produced  in only $0.1s$.  The  same  process occurs  on the neurons in the sequence $ \alpha_2,\alpha_3, ...,\alpha_N$ of the $N$ connected states. Thus, a bunch of  $N$ groups of  $N_{i}$  synchronized neurons for each state $ \alpha$ of the sequence could be expected. 
\par
The neuron firing pulses can be described in terms of functions $F_{\alpha_i}(t)$, where $t$ is the time and  $\alpha_i$ is one of the $N$ states which completely characterizes the firing. The superposition of individual firing $F_{\alpha_i}(t-t_i)$, where $t_i$ is the time origin arbitrarily chosen for each $F_{\alpha_i}(t)$ defines the relevant neurological signal
$\label{Eq1.2}I(t)=\sum_{i=-\infty}^\infty F_{\alpha_i}(t-t_i).$
The  time interval between subsequent events $F_{\alpha_i}(t)$ and $F_{\alpha_{i+1}}(t)$ is indicated by the variable $u_i$, which  depends on the states $\alpha$ through the distribution function  $q_{\alpha_i,\alpha_{i+1}}(u_i)$, accounting for the correlation between the time intervals $u_i$.
\par
The  $\alpha$ states are  represented by an homogeneous Markov chain characterised by the $N \times N$  matrix $\|m\|$  with entries:
\begin{equation}
\label{Eq2.2}
m_{\alpha_i,\alpha_{i+1}} = P({\alpha_{i+1}}=\alpha'|{\alpha_{i}}=\alpha)
\end{equation}
defined as the conditional probability that if the event $i$ is in a state $\alpha $, the event ${i+1}$ will be in the state $\alpha '$, where $\alpha $ and $\alpha '$ are a pair of $N$ states.
Then,  the matrix $\|M(u)\|$   can be built with entries:
\begin{equation}
\label{Eq3.2}
M_{\alpha,\alpha^{'}} (u) = m_{\alpha,\alpha^{'}}q_{\alpha,\alpha^{'}}(u),
\end{equation}
where  $\alpha$ and $\alpha'$ are the states of a pair of successive events of the sequence and  $q_{\alpha,\alpha'}(u)$ the distribution function   of the time interval between the pair.
 The Fourier  transform of $M_{\alpha,\alpha^{'}} (u)$ is defined as:
\begin{equation}
\label{Eq2.4}
M_{\alpha,\alpha^{'}}(\omega) = m_{\alpha,\alpha^{'}} Q_{\alpha,\alpha^{'}} (\omega),
\end{equation}
where:
\begin{equation}
\label{Eq5.2}
Q_{\alpha,\alpha^{'}} (\omega)=\int_0^\infty q_{\alpha,\alpha^{'}} (u) \exp (i \omega u) du ,
\end{equation}
$M_{\alpha,\alpha^{'}}(\omega)$  defines the entries of the correlation matrix $\|M(\omega)\|$ of the Markov process.
\par
The power spectrum of random processes correlated in a Markov chain is given by:
\begin{equation}
\label{Eq2.6}
\Phi(\omega)= \nu  \overline{|S_{\alpha}(\omega)|^2} +2 \nu \Re\sum_{\alpha \alpha^{'}}  S^*_{\alpha}(\omega) S_{\alpha^{'}}(\omega)  p_{\alpha}  \|K(\omega)\|_{\alpha \alpha^{'}}  \, ,
\end{equation}
where $\nu$  is the average number of events per unit time,  $S_{\alpha}(\omega)$  is  of the Fourier transform of  $F_{\alpha}(t)$, with the overline indicating the average over all the pulses in the sequence.  $S_{\alpha}^*(\omega)$ and $S_{\alpha^{'}}(\omega)$  are the Fourier transforms of $F_{\alpha}(t)$ for a pair of states $\alpha$ and $\alpha^{'}$, $p_{\alpha}$ is the fraction of states $\alpha$ in the sequence. $\Re$ means  the real part.  The matrix $\|K(\omega)\|$ is defined as:
\begin{equation}
\label{Eq7.2}
\|K(\omega)\|= \|M(\omega)\| \cdot \left( \|I\| - \|M(\omega)\|\right)^{-1} 
\end{equation}
with $\|M(\omega)\|$ is the correlation matrix  of the Markov process with entries defined by  Eq.~(\ref{Eq2.4}) and $\|I\|$  the identity matrix.
\subsection{Line Power Spectral Density}
In this subsection, we will show how to derive the power spectral density of a single closed loop of Markov-chain correlated pulses. 
When the Markov matrix $\|m\|$ in Eq.~(\ref{Eq2.2}) describes  random events organized in closed loop sequences, the correlation matrix $\|M\|$ yields physically sound line spectra as those observed in brain waves. 
\par
 The lines in the power spectrum   correspond to singularities of Eq.~(\ref{Eq7.2}) when  $\det \left( \|I\| - \|M(\omega)\|\right)=0$, i.e. for  $q_{\alpha,\alpha^{'}} (u)=\delta (u-u_{\alpha,\alpha^{'}})$ where $u_{\alpha,\alpha^{'}}$ is the time interval between consecutive events characterized by the states $\alpha,\alpha^{'}$, then: 
\begin{equation}
Q_{\alpha,\alpha^{'}} (\omega)=\exp(i \omega u_{\alpha,\alpha^{'}})
\end{equation}
and Eq.~(\ref{Eq5.2}) takes the form:
\begin{equation}
M_{\alpha,\alpha^{'}} (\omega)= m_{\alpha,\alpha^{'}} \exp (i \omega u_{\alpha,\alpha^{'}})
\end{equation} 
A closed loop involving $N$ states $\alpha$ can be expressed by a $N\times N $ matrix  $\|M(\omega)\|$ where all the rows are made up of zeroes except one entry equal to $1$. 
with $m_{\alpha,\alpha^{'}} =0 $ except if  $\alpha_{i+1}=\alpha'$ and ${\alpha_{i}}=\alpha $ then  $m_{\alpha,\alpha^{'}} =1$.   The condition   $m_{N-1,N} = m_{N,1} =1   $  ensures closeness of the loops. With the above choices, the relationship $\det \left( \|I\| - \|M(\omega)\|\right)=0$ becomes:
\begin{equation}
\label{Eq2.14}
1-\exp [i \omega (u_{1,2}+ u_{2,3} +...+ u_{N,1} )] =0
\end{equation}
with solutions:
\begin{equation}
\label{Eq2.15}
\omega=\omega_n= n \omega_o =  \frac{2 \pi}{\sum_{i=1}^N u_{i,i+1}} 
\end{equation}
Eq.~(\ref{Eq2.15}) yields the fundamental frequency $\omega_o$ and the harmonics of the periodic component.  $n$ is an integer. $\omega_o$ decreases as $N$ and $u_{i, i+1}$ increase.
The amplitude   of the spectral lines at  $\omega=\omega_n$  is given by: 
\begin{equation}
\label{Eq2.17}
A_n= \frac{2 \pi\nu^2}{N^2}\Re\sum_{\alpha \alpha^{'}}  S^*_{\alpha}(\omega_n) S_{\alpha^{'}}(\omega_n)  C_{\alpha \alpha^{'}} (\omega_n) 
\end{equation}
where $ \nu$  and  $ S^*_{\alpha}(\omega) S_{\alpha^{'}}(\omega) $ have been defined after Eq.~(\ref{Eq2.6}) and $C_{\alpha \alpha^{'}}$ are the entries of the adjoint matrix:
$\|C(\omega_n)\| = \operatorname{adj} \left( \|I\| - \|M(\omega_n)\|\right)$.
The amplitude given by Eq.~(\ref{Eq2.17}) holds only at the singular  frequency values  $\omega=\omega_n$, when  the determinant  is zero. At frequencies $\omega$   different from the singularities $ \omega_n$, the amplitude is equal to zero as expected for strictly periodic functions. 
Thus, the power spectral density of the periodic components writes:
\begin{equation}
\label{Eq15.4}
\Phi_{\ell} (\omega)=   \sum_{n=-\infty}^{\infty}  A_n\; \delta ( \omega -\omega_n )
\end{equation}
The amplitude  $A_n$ has been estimated  for different values of the parameters $\nu$, $N$ and Fourier transform of the single pulse  $S_{\alpha}$ in \cite{Mazzetti2017Harmonic}.
\par
The power spectral density of an arbitrary Markov correlated pulse sequence 
Eq.~(\ref{Eq2.6})  has  been worked out  by averaging over time from ${-\infty}$ to ${\infty}$, by assuming  stationarity. In the case of a single closed loop sequence,   
 the power spectrum of a perfectly periodic signal is obtained, i.e. a line power spectrum taking discrete positive values  at $\ n \, \omega_0 $  and  zero at any other frequency and  no broadband noise is generated. 
The  expression within square brackets  in Eq. (\ref{Eq2.6}) corresponds to the real part of the sum of $ N $ elements of the principal diagonal of the matrix $||K(\omega)||$ multiplied by  $  S_{\alpha}(\omega)S_{\alpha}(\omega)^*$ and by $ p_{\alpha}$,
while the off-diagonal  elements  do not  contribute. The real parts of the diagonal elements  are equal to $-1/2 $, multiplied by $ p_{\alpha}$ $ = $ $1/N  $ and by $ 2$,  give $ - \overline{|S(\omega)|^2}$, that summed to the first term $\overline{|S(\omega)|^2}$ of the same equation, cancel each other.  
The  term $ - \overline{|S(\omega)|^2}$  is the average over all the $N$ states $\alpha$ of the square modulus of products  relative to the two complex conjugate transforms of  a couple of identical impulses symmetric with respect to zero within the two half of the sequence extending from $-\infty $ to $\infty $, while similar products within square brackets in the same equation refers to a couple of different impulses relative to the same state $\alpha$ along the sequence.  When each state $\alpha$ remain identical along the sequence from $-\infty $ to $\infty $ and only their position along the sequence change  according to the conditional probability expressed by the Markov matrix $||m||$, averaging over the statistical ensemble does not change the above conclusions. The line spectrum, generated by a periodic function, remain unchanged by averaging along the sequence.
\subsection{Broadband Power Spectral Density}
In this subsection,  the Markov matrix approach is extended to an arbitrary number $N_i$ of synchronized closed loops of firing neurons. Under the assumption of fluctuations of $N_i$ the expression of the  power spectrum containing the broadband components is derived.
The syncronized firing of  neurons belonging to the same state $\alpha$  will be described by the superposition of individual firing functions in terms of the  $F_{\alpha}(t)=\sum_{i=1}^{N_i}F_{i,\alpha}(t)= N_i \langle {F_{i,\alpha}(t)}\rangle$, 
where the angle brackets $\langle {\,\,}\rangle$ refers to the average over the ensemble of the $N_i$ loops.  Owing to the large number $N_i$ of  loops  and the imperfect synchronization of the elementary impulses belonging to the same $\alpha$ state, the firing events  are described as Gaussian functions, hence $\langle {F_{i,\alpha}(t)}\rangle$ can be written as: 
\begin{equation}
\label{Eq3.2}
\langle F_{i,\alpha}(t)\rangle =\frac{\langle {A_{\alpha}}\rangle}{\sigma_{\alpha} \sqrt{2 \pi}} \exp \left(-\frac{t^2}{2\sigma_{\alpha}^2}\right) \quad , 
\end{equation}
with variance $\sigma_{\alpha}$, amplitude  $ \langle {A_{\alpha}}\rangle = \int_{-\infty}^\infty {\langle F_{i,\alpha}(t)\rangle dt}$ and the  time origin of the impulses taken at the maximum of the Gaussian function relative to every state $\alpha$.
The Fourier transform of the function $F_{\alpha}(t)$ can be written as $S_{\alpha}(\omega)= N_i \langle S_{i,\alpha}(\omega) \rangle$ where:
\begin{equation}
\label{Eq3.3}
 \langle S_{i,\alpha}(\omega) \rangle = \langle {A_{\alpha}}\rangle  \exp\left(-\frac{\omega^2 \sigma_{\alpha}^2}{2}\right) \quad.
\end{equation}
\par
When the fluctuations of the number $ N_i $ of synchronized loops  are taken into account in the general expression of the power spectal density, 
the quantity  $S_{\alpha}(\omega)S_{\alpha}(\omega)^*$ writes $(\overline{ N_i})^2 \overline{(<| S_{i,\alpha}(\omega)|>)^2 } $. The first term in the same equation is $\overline {(N_i^2 )}$
$\overline {(<|S_{i,\alpha}(\omega)|>)^2}$. 
Hence, the power spectrum writes:
\begin{equation}
\label{Eq3.4}
\Phi (\omega)=\nu \left(\overline{N_i^2}-\overline{N_i}^2\right) \overline {(\langle|S_{i,\alpha}(\omega)|\rangle)^2} 
\end{equation}
The quantity $ \overline{N_i^2}-\overline{N_i}^2$,  yielding the fluctuations of $N_i$ around its average value $ \overline N_i$,  can be estimated by assuming that $N_i$ is a stationary random variable described by a normalized  Gaussian probability function $P(N_i)$:
\begin{equation}
\label{Eq3.5}
P(N_i)=\frac{1}{\sigma_{N_i}\sqrt{2 \pi}}\cdot \exp\frac{-(N_i-\overline{N_i})^2}{2\sigma_{N_i}^2} \quad,
\end{equation}
with 
$ (\overline{N_i})^2 = \left(\int_{-\infty}^\infty \left( N_i\cdot P(N_i) \right) dN_i\right)^2$   and $\overline {(N_i^2 )} =  \int_{-\infty}^\infty \left(N_i^2 \cdot P(N_i) \right) dN_i=(\overline{N_i})^2+\sigma_{N_i}^2 $ and the variance is $\sigma_{N_i}$. Hence  Eq.~(\ref{Eq3.4}) writes:
\begin{equation}
\label{Eq3.8}
\Phi (\omega)=\nu \sigma_{N_i}^2 \overline {\langle|S_{i,\alpha}(\omega)|\rangle^2} \quad ,
\end{equation}
which yields the broadband component of the noise power spectrum, whose intensity has been estimated for normally distributed fluctuations of the number of syncronysed loops. The amplitude of the broadband component depends on the fluctuations of $N_i$ through the variance $\sigma_{N_i}$. 
\par
The width of the Gaussian $\sigma_{\alpha}$  for each state $\alpha$  in Eq.(\ref{Eq3.3}) is related to the cut-off frequency of the power spectrum and thus, to the cut-off of the highest harmonic frequencies of the periodic components. 
If the pulses are Gaussian with Fourier transform given by Eq.~(\ref{Eq3.3}),
the cut-off angular frequency  may be assumed to be $\omega_{c}=1/\overline\sigma_{\alpha}$, giving a reduction factor of $ 1/e \approx 0.36 $.
For instance, $\overline\sigma_{\alpha}=2 \cdot 10^{-3} s$ yields a cut off frequency $ \omega_{c}=500$ $rad \, s^{-1}$ and  $f_{c} =79.5 Hz $, that for an  \textit{alpha}  wave of $ 8 Hz $  would allow about $10$ harmonics to stand  in the power spectrum up to $ 80 Hz $ with only a slight reduction, while for a wave of $ 20 Hz $ a reduction of the amplitude would occur after the $4^{th}$ harmonic. 
At frequencies lower than the cut-off $ \omega_{c}$, number and amplitude of the harmonics  merely depend  on the states characterizing the neurons forming the loops.
\par
The main effect of the fluctuations of $N_i$ is to give rise to a mixed power spectrum.  The effect of this noise adds to the signal, which now includes  spectral lines at  the angular frequencies $\ n \cdot \omega_0 $ and a continuous broadband noise. Fluctuations of the parameters intrisic to a single loop give rise to the changes in the synchronization with the other loops by changing the number $N_i$.   As observed in empirical EEG records, amplitude and waveform vary almost at every period of the detected signal. According to our model, this change may be due to fluctuations of the number of synchronized loops $N_i$  and onset of synchronization  with slightly different characteristics.  
\section{Discussion}
In this section, the power spectrum generated by the  synchronized closed loop sequences of Markov correlated firing neurons  will be confronted with typical features observed in  EEG and MEG measurements. As discussed in Sect. 2, the sequences of electric and magnetic impulses received by the sensors on the scalp at every cycle  correspond to $N$ groups  of $N_i$ syncronized impulses generated by the firing of a single neuron in each loop. The average time intervals $u$  between subsequent neuron firings along the loop multiplied by $N$  yields the loop duration and its reciprocal the lowest frequency component of the mixed power spectrum.  
In principle $N_i$ could be  estimated by considering the intensity of either the elementary electric or the magnetic impulses, which are  almost simultaneously  produced by a single firing neuron in each sequence.  Better estimates are obtained by using the magnetic component, not attenuated by the cerebral matter and the scalp, contrarily to the electrical components of the signal \cite{Izhikevich2007Dynamical,Delorme2012Independent,Vorwerk2014A}. The magnetic  and  electric field components $\vec{B}$ and $ \vec{E} $  generated by the firing of a single neuron obey the Maxwell equation $\curl \,\vec{B} =\mu_0 \vec{J} + \epsilon_0\mu_0 \,\, {\delta \vec{E}}/{\delta t }$, with   $ \vec{J} $  and  $\mu_0 \epsilon_0 \,\,{\delta \vec{E}}/{\delta t }$   the conduction and   displacement  current density. The conduction charges in the axon move at  speed  ranging between $0.5 m s^{-1}$ and $ 5.0 m s^{-1}$,  thus much faster than the charges moving in the outer regions.
Therefore, the electric field  in the axon is partly screened by the conductive cerebral matter.
\par
 The magnetic field generated by the conduction charge inside the axon can be estimated by using the relationship valid for metallic conductors:
\begin{equation*} 
\label{Eq4.2}
\vec{B}(t) = \frac {\mu_0}{4\pi} I(t)   \, \Delta \ell  \, \frac{\vec{u}_{A}\,\times \,\vec{u}_{D}}{r_s^2} \hspace{3pt},
\end{equation*}
with $I(t)$ the current intensity,  $r_s$ the distance between the midpoint of the axon and the point  where $ \vec{B}(t)$ is measured,    $\Delta \ell $  the length of the axon,  which, for neurons connected in the same area of the brain, ranges  between  $50\mu m $ and $ 200  \mu m $.  The unit vector $\vec{u}_{A} $ and $\vec{u}_{D}$ indicate respectively the       directions of $\Delta \ell $ and  $r_s$.  The amplitude of the magnetic field  $\vec{B}(t)$  writes:
\begin {equation}
\label{Eq4.3}
{B}(t) = \frac {\mu_0}{4\pi} \, I(t) \,  \Delta \ell \ \frac {\sin \theta}{r_s^2}\hspace{10pt} ,
\end{equation}
where $\theta$ is the angle between $ \vec{u}_{A}$ and $\vec{u}_{D} $.
  $I(t)$ can be estimated by considering the charge transferred by the firing process to nearby neurons. When the neuron receives positive inputs from its dendrites   over a short time interval,  its resting membrane potential increases from about   $ - 75 mV $ to a critical value of about $ - 45 mV $. At this point,  a  positive charge $Q$ enters the soma from the ionic channels  making the membrane potential slightly positive a process lasting  about $1 ms$. During a subsequent time interval with approximately the same duration ($1ms$) the excess positive charge $Q$  is ejected through its axon, restoring the membrane potential to its resting value $-75 mV $  after a small over-shut.  As a good approximation, the ejection of this charge corresponds to a variation of the membrane potential of about $ 100 mV $.
By approximating the soma as a sphere of radius $R$ with uniform inner charge density $ \rho $, the electric field  can be written as 
$E(r)=    { \rho \frac {4}{3} \pi r^3 }/{4 \pi \epsilon_0 r^2} =  {\rho r}/{3 \epsilon_0}\,$ at $ r\le R$ with
$ r $  the  distance from the center of the sphere. The membrane potential $ V_{m} $ writes:
\begin {equation}
\label{Eq4.5}
V_{m} = \int_{0}^R E(r) dr = \frac {\rho}{3 \epsilon_0}  \int_{0}^R r dr = \frac {\rho}{6 \epsilon_0} R^2 = \frac {Q}{8\pi \epsilon_0 R }
\end{equation}
where $ Q=\rho\tfrac{4} {3}\pi R^{3}$ is the total charge within the sphere.
A membrane potential of about $ V_{m} = 100 mV $ and a soma radius of  about $ R = 25.0\,\, \mu m $ result in an ejected charge 
$Q = 8\pi \epsilon_0 R  V_{m} =5.55 \cdot 10^{-16} \, C $. If the ejected charge crosses the axon in about $1 ms$, the current impulse $ I(t) $ can be described by a Gaussian function of time with variance $ \,\sigma = 1 ms $:
\begin {equation}
\label{Eq4.7}
I(t) =  \frac{Q} {\sigma \sqrt{2\pi}} \exp\left(\frac{ t^2}{2\,\sigma^{2}}\right)
\end{equation}
with the  condition $ \int_{-\infty}^\infty  I(t)dt = Q $.
The current $I(t)$ can be introduced in the magnetic field (Eq.~(\ref{Eq4.3})), i.e.:
\begin {equation}
\label{Eq4.9}
{B}(t) = \frac {\mu_0}{4\pi} \frac{Q} {\sigma \sqrt{2\pi}}\exp \left (\frac{ t^2}{2\,\sigma^{2}}\right) \Delta \ell  \frac {\sin(\theta)}{r_s ^2} \hspace{5pt}.
\end{equation}
The  maximum value  of the magnetic field amplitude is achieved at $t=0$ with $\sin\theta=1 $. This situation, where $\sin(\theta)=1 $, occurs when the axon of the firing neuron is tangent to the surface of the skull in the point where  the magnetic sensor is placed. In the cortical area of the brain this may happen when the firing neuron is in one of the numerous regions called gyri \cite{Jiang2021Fundamental,Wang2022Modeling}.
Then by using an average value of $\Delta\ell = 100\,\,\mu m$ and of the distance $ r_s = 2 \cdot 10^{-2}\,\, m $ of the magnetic sensor detecting the impulse, an average peak intensity
${B}(0) \approxeq 0.5\cdot \,10^{-20} \, T$ is obtained.   To generate magnetic impulses of the order of $ 10^{-13} \,T $, value generally found in MEG and EEG, about $N_i=10^7$ synchronized firing neurons would  be needed \footnote {The value  $ 10^{-13} \,T $ is close to the sensitivity limit of the  SQUID sensors (about $ 10^{-15} T $).}
\par
While it is generally accepted that the magnetic signal  generated  by  firing neurons (MEG) is due to the  ejected excess charge within the axon, the  origin of the electric signal  (EEG)   is more controversial.  A common assumption is that the signal is generated in correspondence of the chemical synapses  connecting the firing neuron to the dendrites of numerous postsynaptic ones. The charge emitted from the axon of the firing neuron is split  in hundreds or thousands fractions which create ionic currents external to the synapses, then detected by the sensors of the EEG setup. The synchronized firing of a large population of neurons and the role played  by the excitatory and inhibitory synapses yield a complex situation of small current impulses, where spontaneous oscillations can be generated under suitable assumptions \cite{Hutt2010Oscillatory,Hascemi2015How}.
An alternative interpretation  relates the electric signals   to the transient potential impulses radially emitted externally to the soma of the firing neurons, due to the rapid variations of the external electric field generated by the charge variations inside the neuron. Within this description, the potential impulse peak value, detected by an electrode internal to the soma during firing, is assumed of the order of $100 mV$, the electrode external to the soma peak value of about $0.1 mV$, the distance of the second electrode from the membrane of the neuron of the order of $ 200 \mu m $.  In order to evaluate the intensity of the electric potential impulse,  we consider again the neuron with the sferical soma of radius $ R = 25\mu m $ and the electric field $E(r)$ at $ r < R $ used above for the calculation of the magnetic impulse.  If the conductive medium within the cranial bone is neglected, the electric field $E(r)$ at $ r > R $ is given by $E(r)= {Q}/{4 \pi \epsilon_0 r^2} \, $
where $Q$ is the whole charge within the soma.  
The electric potential, at distance $ r >R $ from the center of the sphere,  is given by $ V =    {Q }/{4 \pi \epsilon_0 r} $, with $V=0$   as $ r = \infty $.
The field $E(r)$   in the presence of conductive liquids, as it is the case for the brain, in stationary condition would be zero. A charge layer equal to the internal charge but with opposite sign, would be attracted and surround the membrane of the soma. This charge with spherical symmetry cancels out the field for $ r > R $.
When a fast transient process occurs to the charge within the soma, as it is the case during the firing of the neuron, and the conductive liquid within the cranial bone contains positive and negative ions, the screening of the charge inside the soma is expected to occur only in part, which justifies the impulse \cite{Hutt2010Oscillatory}. The amplitude of this impulse, compared to the amplitude expected from Eq.~(\ref{Eq4.5}), allows to evaluate the effective charge $ Q_{eff}$ at the center of the sphere, and thus the amplitude of the impulse at an arbitrary distance $ r >R $.
By assuming a distance $ r$  from the center of the sphere of the order of $ 200 \mu m $ and $ Q = 5.55 \cdot 10^{-16} \;C$,   Eq.~(\ref{Eq4.5}) gives $ V = 22.28 mV$,  exceeding the typical  measured value $ 0.1 mV$.  This result could imply that the charge internal to the soma is screened by an external opposite charge which reduces the $Q$ value to $ Q_{eff}$. A value of  $ Q_{eff}$ of the order of $10^{-18} C $, i.e. two orders of magnitude lower than $Q$  would yield a potential value of the order of $ 0.1 mV $ at a distance of about $200 \mu m $  and a potential value of the order of $ 10^{-6} V$ at a distance of about  $  10^{-2} m $. 
It should be also  considered that the electrometer input impedance is close to $\infty$. The impulse amplitude is expected  to be  reduced by the cranial bone and by the input impedance of the operational amplifiers connected to the electric sensors of the EEG set-up, with a gain inversely proportional to the input resistance, that must be  kept low  but large enough to drive the signal recorder. By keeping into account all these effect a smaller value of the potential could be obtained.
\section{Concluding remarks}
A mathematical framework, leading to the unified description of the discrete and continuous power spectral components of the  EEG and MEG signals, has been proposed.
 The model is based on the assumption that  the complex network of interacting neurons spontaneously form closed loops chains of $N$  firing neurons  giving rise to repetitive emission of   electric and magnetic impulses.  
By describing  the electric and magnetic impulses generated by the firing of a neuron as  Gaussians of given amplitude and width, the line power spectrum  of this periodic function, fundamental and harmonic components, are shown to depend on the distributions of the amplitude and variance of the emitted impulses and of the time intervals between their emission, which are expression of the $N$ states $\alpha$  of the sequence.  A general matrix equation gives the line power spectrum.  
\par
Furthermore, by considering the fluctuations of the number $N_i$ of synchronized firing neurons belonging to different  closed loop chains, a broadband component emerges in the power spectrum.  The syncronized pulses are represented as  Gaussian time functions whose amplitude and variance depend on the degree of syncronization and  simultaneousness of the firing of the neurons  belonging to different single loops but in the same state $\alpha$. T
\par
 As discussed in Sect.3,  the fluctuations of the number of synchronized loops $N_i$ may change the amplitude of the signal and  its waveform which is determined by the harmonic components generated by the loops. It is worthy of note that, even if the fluctuations change the waveform of the signal, the frequency of the wave remains  unchanged, as the frequency of the wave depends on the duration of the loop $\sum_i u_i$ which is unchanged by the fluctuation of $N_i$.
An additional effect  of the presence of the broadband component concerns the onset of a cut-off in the line power spectrum, which is related to the width of the electric impulses  represented by the variance of the  Gaussian time functions.  Larger variance implies lower  cut-off frequency and smaller number of harmonic components, filtering out a signal closer and closer to the pure sinusoid. 

The main  features explained by the proposed model are summarized here below.
The broad range of discrete frequencies observed in the EEG and MEG signals is obtained by considering  different duration of closed loops sequences of firing neurons.  By assuming an average firing interval between successive neurons of $  5\cdot 10^{-3}$ s, sequences ranging  from few hundred neurons to only few neurons cover the range from $0.5 Hz$ (lowest frequency of \textit{delta} waves) to  50 Hz (highest frequency of \textit{gamma} waves). The presence of several harmonic components of the fundamental sinusoidal wave in the graph of EEG is also accounted for, particularly at low frequency, as the \textit{delta} waves, where a pure sinusoid is never observed. High frequency waves are less distorted since harmonics stand at frequencies multiple of the fundamental one, and the cut-off  frequency of the detected power spectrum cuts the harmonics exceeding that frequency. In a few particular cases all harmonics are cut-off, or strongly reduced, giving a sinusoidal, or nearly sinusoidal, wave.  Part of the distortion can be due to the broadband noise created by the fluctuation of the number of $N_i$. One of the effects discussed above, and observed in almost all EEG graphs, is the continuous change of the waveform, practically at every period, of the received signal produced by the loops. This change, which is associated to a heavy distortion of the fundamental wave, is due to the presence of several harmonic waves within the periodic signal, and it is enough a change of the amplitude or the phase of one or few of these harmonics  during the fluctuation of $ N_i $, to change the waveform over the period. A signal constituted of a fundamental wave and many harmonics is just what is expected from a sequence of heterogeneous electric impulses characterized by different states $\alpha$, as  considered in the present paper. 


\end{document}